\pdfoutput=1

\documentclass[]{aastex631}
\usepackage{amsmath}\usepackage{amssymb}\usepackage{natbib}\usepackage{CJKutf8}\usepackage{orcidlink}

\begin{document}
\begin{CJK}{UTF8}{gbsn}
\title{A Generic Model for Persistent Radio Source around Fast Radio Bursts}

\author{Yushan Chen}
\affiliation{School of Physics and Materials Science, Guangzhou University, Guangzhou 510006, China}

\author{Hao Tong~\orcidlink{0000-0001-7120-4076}}
\affiliation{School of Physics and Materials Science, Guangzhou University, Guangzhou 510006, China}

\correspondingauthor{Hao Tong}
\email{tonghao@gzhu.edu.cn}


\begin{abstract}

The repeated fast radio burst FRB 121102A and FRB 190520B has been reported, along with a spatially coincident, compact, persistent radio emission.
In this paper, we present a parameterized one-zone model, with a basic scenario that a relativistic magnetized wind from the pulsar sweeps up the surroundings, e.g. freely expanding supernova ejecta, giving rise to a power-law distribution of electron filled between the forward shock and the termination shock.
We show that via appropriate adjustment of the model parameters, we can obtain the synchrotron radio emission properties from the one-zone model bright enough to account for observation, simply and analytically fitting the observed spectra well. Through dynamical evolution of the model, we can also obtain time-varying of relevant properties. This parameterized model does not depend on concrete physical models such as central engine, instead we can constraint physical model via comparison between parameters and observation, indicating the information about the central engine and surroundings.
We also discuss the synchrotron self-Compton emission in our scenario in the end, but find no clue on the counterparts at other waveband.

\end{abstract}



\section{Introduction} \label{sec:intro}

Fast radio bursts (FRBs) are short pulses of coherent radio
emission of unknown physical origin lasting less than a few milliseconds \citep{2007Sci...318..777L,2012MNRAS.425L..71K,2013Sci...341...53T,2014ApJ...790..101S,2015ApJ...799L...5R,2016Sci...354.1249R,2016MNRAS.460L..30C,2016Natur.531..202S,2016PASA...33...45P,2017Natur.541...58C,2019ARA&A..57..417C,2019A&ARv..27....4P,2022A&ARv..30....2P}. Most of them has large dispersion measures (DMs; DM $\approx 300-2000 
 \ \rm pc \ cm^{-3} $), well above the
contribution from the Milky Way and thus implicating an extragalactic origin. The cosmological distance of FRB 121102A was confirmed \citep{2014ApJ...790..101S,2016Natur.531..202S} and its subsequent
localization \cite{2017Natur.541...58C} to a dwarf star-forming galaxy at a redshift of $z \approx 0.1927$ \citep{2017ApJ...834L...7T} further  confirms a cosmological origin for this FRB. 

More importantly, radio  interferometric localization of FRB 1211012A revealed a
compact (size $< 0.7 \rm \rm pc$) luminous ($\nu L_\nu \sim 10^{39} \ \rm erg \ s^{-1}$) persistent radio source (PRS) coincident to within $\leq40 \ \rm pc$ of the
FRB location \citep{2017ApJ...834L...8M,2022arXiv220100999C}. The second PRS counterpart of FRB 190520B was confirmed later and located in a dwarf galaxy ($\leq 20 \rm \ mas$) similar to that of  FRB 121102A \citep{2022Natur.606..873N,2023arXiv230812801B}. 
Although the origin of PRS remains unclear, these two PRSs show common in spectra properties \citep{2023arXiv230716355Z} and thus the comparison of bursts arises between FRB 121102A and FRB 190520B \citep{2023MNRAS.522.5600L}.
Another important clue to these two FRBs comes from their enormous rotation measure (RM), RM $\sim 10^5 \ \rm rad \ m^{-2}$ \citep{2018Natur.553..182M,2022NatAs...6..393N} and large dispersion measure (DM) \citep{2021Natur.598..267L,2022Natur.606..873N}, which indicates a dense and strong-magnetized environment.
There has been proposed that the large DM and scattering of FRB 190520B is dominated by the host galaxy, making the bursts a probe into the circumstances, e.g. the plasma and turbulence \citep{2022ApJ...931...87O,2023MNRAS.519..821O}.
Since counterparts like PRS can unveil insights of FRB itself, search for other counterparts in different waveband (e.g. optical) has been raised \citep{2023arXiv230311967V,2023arXiv230312598V,2022ApJ...929..139L,2022MNRAS.515.1365C}, yet no optical counterpart of FRB 190520B was confirmed \citep{2022ApJ...931..109N}.
With discovery of an FRB in a clean environment \citep{2023arXiv230414671F} and suggestion that there is no obvious evidence for repeaters or non-repeaters to be preferentially associated with PRS \citep{2022ApJ...927...55L},
this may indicate that FRB and PRS are two aspects of the central engine rather than the same origin, as pointed out that the polarization of PRS differs from that of FRB in \cite{2023arXiv230812801B}. 

Though dozens of models have been proposed for FRBs, most are ruled out by a repeating, cosmological source like FRB 121102A. Two types of central engine for FRBs are discussed most. First, for a rotationally powered model \citep{2016MNRAS.457..232C,2016MNRAS.458L..19C,2016MNRAS.462..941L,2017ApJ...839L...3K,2017ApJ...841...14M,2021ApJ...923L..17Z}, FRBs from a millisecond
magnetar are scaled-up versions of super-giant pulses from the
Crab pulsar. Second, for a magnetically powered model \citep{2013arXiv1307.4924P,2014MNRAS.442L...9L,2014ApJ...797...70K,2016ApJ...826..226K,2017MNRAS.468.2726K,2017ApJ...839L...3K,2017ApJ...841...14M,2018ApJ...868L...4M,2021ApJ...923L..17Z}, FRBs may
arise from the unexpected release of magnetic energy in the magnetar’s interior or
electrostatic energy  \citep{2017MNRAS.469L..39K}, similar to the giant flare model of Galactic magnetars. By discovery of FRB 121102A and its PRS, particularly its small size and magnetized environment, such a model was proposed that the spatially coincident PRS could be understood as emission from a compact magnetized nebula surrounding the young (decades to
centuries old) neutron star, embedded behind the expanding supernova ejecta shell or ambient intersteller medium \citep{2017ApJ...839L...3K,2017ApJ...838L...7D,2017ApJ...841...14M,2018ApJ...868L...4M,2019ApJ...885..149Y,2020ApJ...896...71L,2021ApJ...923L..17Z}. As comparison was made for FRB and gamma ray burst \citep{2020Natur.587...45Z}, the surrounding persistent radio emission is similar to the afterglow of gamma ray burst. Most likely, the nebula is powered by nearly continual energy release from the
magnetar, likely during the same sporadic flaring events responsible for the repeated radio bursts \citep{2016ApJ...833..261B}.
Meanwhile, the PRS could be driven by different energy budget, i.e. the rotational or the magnetic power or even the FRB bursts \citep{2020ApJ...896...71L}.

In our paper, we present a parameterized one-zone model of PRS with energy injection from central engine, and show how to fit the observed spectra and constraint the parameters through observation regardless of energy budget and physical model. More specifically, we consider the situation of a electron-proton pulsar wind nebulae (PWN) embedded in a free expanding supernova remnant (SNR), and we also compare other situation via adjusting relative parameters. 

In section \ref{sec:Parameter estimation}, we make an estimation on the parameters.
In section \ref{subsec:radiation}, we analyse  
radio emission properties of one-zone model to fit the observed spectrum energy distribution of PRSs of FRB 121102A and FRB 190520B (denoted as PRS 121102 and PRS 190520 as follows). We also present the dynamical evolution of the model for a specific case in section \ref{subsec:dynamical}. 
In section \ref{sec:results}, we constraint the model parameters and discuss the energy injection.
In section \ref{sec:discussion}, we
present our discussion about the other properties of one-zone model. 

\section{Analytial one-zone model}
\label{sec:theory}
\subsection{Parameter estimation} \label{sec:Parameter estimation}

One possible source engine for powering both FRB and its PRS may be magnetar, earliest introduced by \cite{2010vaoa.conf..129P}, then \cite{2014MNRAS.442L...9L} presented the interaction between magnetar flare and the ambient wind as a mechanism to generate FRBs. \cite{2017ApJ...838L...7D} and \cite{2018ApJ...868L...4M} pointed out that magnetic energy from magnetar could be responsible for PRS 121102,
\begin{equation}
    E_{\rm \rm mag}\sim{B^2R^3}\sim{10^{50} B^2_{\rm 16}R^3_6 \rm \ erg},
\end{equation}
Where $B\sim 10^{16} B_{\rm 16} \rm \  G$ is the surface field strength of magnetar, $R\sim 10 R_6 \rm \ km$ is the typical neutron star radius. On the other hand, the rotation energy,
\begin{equation}
    E_{\rm \rm rot}\sim \frac{1}{2}I\Omega^2\sim 10^{46} I_{\rm 45} P^{-2} \rm \ erg,
\end{equation}
can also be important at the early time evolution, but neglectable at late time due to the rapid rate of magnetic braking. The magnetic energy budget gives rise to an active time of the PRS, when using the magnetic energy, for PRS 121102 $\nu L_\nu \sim 3.4\times10^{38} \rm \  erg \ s^{-1}$ at 1.7GHz \citep{2017ApJ...834L...8M},
\begin{equation}
    t_{\rm \rm active}\leq \frac{E_{\rm \rm mag}}{\nu L_\nu } \approx 9.3\times 10^3 B_{\rm 16}^2 R_6^3 \rm \ yr,
\end{equation}
for PRS 190520 $\nu L_\nu \sim10^{39} \rm \ erg \ s^{-1}$ at 3GHz \citep{2022Natur.606..873N},
\begin{equation}
    t_{\rm \rm active}\leq \frac{E_{\rm \rm mag}}{\nu L_\nu } \approx 3.6\times 10^3 B_{\rm 16}^2 R_6^3 \rm \ yr,
\end{equation}
under the assumption of isotropic radiation.

Another limit on active time for PRS is the size $\sim 0.66 \ \rm pc$ for PRS 121102 according to \cite{2017ApJ...834L...8M} and the size $\sim 4.5 \ \rm pc$ for PRS 190520 according to \cite{2023arXiv230812801B}. We use the velocity of pulsar wind nebulae(PWN) as a constant $\sim 1.18\times 10^{16} \rm \ cm \ yr^{-1}$ \citep{2020Vink}, and put a limit on active time, for PRS 121102
\begin{equation}
    t_{\rm active}\leq 1.7\times 10^2 \rm \ yr,
\end{equation}
for PRS 190520
\begin{equation}
    t_{\rm active}\leq 1.2\times 10^3 \rm \ yr.
\end{equation}
Note that PWN could expand rapidly in its early epoch instead of expanding uniformly.
An additional lower limit $t_{\rm active}\ge 10\sim100 \rm \ yr$ 
for PRS 121102 follows from the requirement that supernova ejecta be transparent to free-free absorption at 1GHz and not to overproduce constraints on time derivative on DM \citep{2018ApJ...868L...4M}.
Typical magnetic energy budget and size limit show that both PRSs should be active and emit at their early age.

Next we estimate the typical value of Lorentz factor of the electrons of the system.
From \cite{1979R&L} we have
\begin{equation}
    \nu_{\rm obs}=\frac{3q_{\rm e}B}{4\pi m_{\rm e} c} \gamma^2,
\end{equation}
and we obtain $\gamma=\frac{20}{\sqrt{B}}$ for PRS 121102. For synchrotron radiation, power emitted by single electron,
\begin{equation}
    P=\frac{4}{3} \sigma_T c \gamma^2 U_{\rm B}\sim 1.0\times 10^{-13}\rm \ erg \ s^{-1},
\end{equation}
where $U_{\rm B}=\frac{B^2}{8\pi}$ is the magnetic energy density and we assume magnetic field $B\sim 0.24\rm \ G$ \citep{2018ApJ...868L...4M}. Then we obtain the total number and total energy of electrons of the system,
\begin{equation}
    N_{\rm e}=\frac{L_{\rm w}}{P}\sim 3.3\times 10^{51},
\end{equation}
where $L_{\rm w}=\nu L_\nu$ is the luminosity, and
\begin{equation}
    E_{\rm e}=N_{\rm e} \gamma m_{\rm e} c^2\sim 1.1\times10^{47} \rm \ erg,
\end{equation}
which means only a small fraction of energy being injected into the electrons compared to the energy budget.

\subsection{Synchrotron Radiation for 
 One-zone Model}\label{subsec:radiation}
For high brightness temperature sources like PRSs, e.g. $T_{\rm B} \ge 5\times10^7\rm \ K$ at 5 GHz for PRS 121102\citep{2017ApJ...834L...8M},
synchrotron radiation is the most possible mechanism to be account for PRSs \citep{2017ApJ...838L...7D,2017ApJ...839L...3K,2017ApJ...841...14M,2018ApJ...868L...4M,2019ApJ...885..149Y,2020ApJ...896...71L,2021ApJ...923L..17Z}.In this subsection, we consider a power-law synchrotron emission with an energy injection to explain the spectra of PRSs, which is similar to explain a plateau in the light curve of an early afterglow of gamma ray bursts with synchrotron fireball model \citep{1998ApJ...497L..17S}.
First we consider an energy injection,
\begin{equation}
    E_{\rm inj}= 10^{50} E_{\rm 50} \rm \ erg,
\end{equation}
and a size,
\begin{equation}
    R_{\rm PWN}= 10^{17} R_{\rm 17} \rm \ cm,
\end{equation}
then we obtain the energy density,
\begin{equation}
    U=\frac{E_{\rm inj}}{\frac{4}{3}\pi R_{\rm PWN}^3}=2.3\times 10^{-2} E_{\rm 50} R_{\rm 17}^{-3}\rm \ erg \ cm^{-3}.
\end{equation}
Next we discuss the synchrotron radiation from the PRS. We introduce $\epsilon_{\rm e} U$ and $\epsilon_{\rm B} U$ as the enenrgy density of electron and magnetic field, noting that $\epsilon_{\rm e}+\epsilon_{\rm B} \neq 1$ since there are other energy depletion in our scenario. We assume a power-law distribution of electron, $\frac{dn_{\rm e}}{d\gamma}=K_{\rm e}\gamma^{-p}$, which has upper and lower limits of $\gamma$. From
\begin{equation}
    \epsilon_{\rm e} U=\int_{\rm \gamma_{\rm min}}^{\gamma_{\rm max}}\frac{dn_{\rm e}}{d\gamma} \gamma m_{\rm e} c^2 d\gamma,
\end{equation}
where $m_{\rm e}$ is the electron mass and $c$ is the speed of light, we obtain
\begin{equation}
    K_{\rm e}=\frac{(2-p)\epsilon_{\rm e} U}{m_{\rm e} c^2 (\gamma_{\rm max}^{2-p}-\gamma_{\rm min}^{2-p})}
\end{equation}
and the total electron number
\begin{equation}
    N_{\rm e}=\frac{4\pi R_{\rm PWN}^3 K_{\rm e}}{3(p-1)\gamma_{\rm min}^{p-1}}.
\end{equation}
The synchrotron emission spectrum depends on three break frequencies. The first break frequency is the synchrotron cooling
frequency at which an electron with the cooling Lorentz factor $\gamma_{\rm c}$ loses its energy in a dynamical time $t$. Using $\gamma_{\rm c}=\frac{6\pi m_{\rm e} c}{\sigma_T B^2 t}$ \citep{1998ApJ...497L..17S}, where $\sigma_T$ is the Thomson cross-section and $B=(8\pi \epsilon_{\rm B} U)^{\frac{1}{2}}$  is
the magnetic field strength in the PWN, the synchrotron cooling frequency is calculated by
\begin{equation}
    \nu_{\rm c}=\gamma_{\rm c}^2 \frac{q_{\rm e} B}{2\pi m_{\rm e} c},
\end{equation}
where $q_{\rm e}$ is the electron charge. The second break frequency is the typical synchrotron frequency which an electron with $\gamma_{\rm min}$ radiates,
\begin{equation}
    \nu_{\rm m}=\gamma_{\rm min}^2 \frac{q_{\rm e} B}{2\pi m_{\rm e} c}.
\end{equation}
The third break frequency is the synchrotron self-absorption frequency \citep{2003MNRAS.342.1131W},
\begin{equation}
\nu_{\rm a}=(\frac{c_2 q_{\rm e} K_{\rm e} R_{\rm PWN}}{B})^\frac{2}{p+4} \frac{q_{\rm e} B}{2\pi m_{\rm e} c},
\end{equation}
where
\begin{equation}
    c_2=\frac{32\pi}{9} 2^{\frac{2}{p}}\frac{\sqrt{3}}{16} \Gamma(\frac{3p+2}{12}) \Gamma(\frac{3p+10}{12}) (p+\frac{10}{3})
\end{equation}
and $\Gamma$ is the $\Gamma$ function. The maximum peak luminosity is calculated by
\begin{equation}
    L_{\rm \nu,max}=N_{\rm e} \frac{m_{\rm e} c^2 \sigma_T}{3q_{\rm e}} B.
\end{equation}
Thus, the synchrotron emission luminosity at any frequency $\nu$ is given by 
\begin{equation}\centering
    L_{\rm \nu}=\begin{cases}L_{\rm \nu,max}(\frac{\nu_{\rm a}}{\nu_{\rm m}})^{-\frac{p-1}{2}}(\frac{\nu}{\nu_{\rm a}})^{\frac{5}{2}}, \ \nu<\nu_{\rm a},
    \\L_{\rm \nu,max}(\frac{\nu}{\nu_{\rm m}})^{-\frac{p-1}{2}}, \ \nu_{\rm a}<\nu<\nu_{\rm c},
    \\L_{\rm \nu,max}(\frac{\nu_{\rm c}}{\nu_{\rm m}})^{-\frac{p-1}{2}}(\frac{\nu}{\nu_{\rm c}})^{-\frac{p}{2}}, \ \nu\ge \nu_{\rm c}.
        
    \end{cases}
\end{equation}
Here we focus on the hard electron spectrum when $\nu<\nu_{\rm c}$.

\begin{figure}[ht!]
\centering
\includegraphics[scale=1]{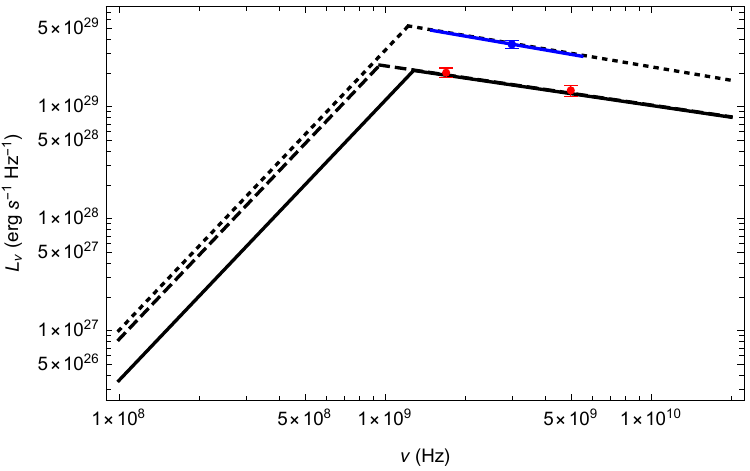}
\caption{\textbf{Here we use the seven parameters, the energy injection $E_{\rm 50}$, the size $R_{\rm 17}$, $\epsilon_{\rm e}, \epsilon_{\rm B}, \gamma_{\rm max}, \gamma_{\rm min}, p$, as detailed in subsection \ref{subsec:radiation}}. For PRS 121102, $E_{\rm 50}=3, R_{\rm 17}=1.5, \epsilon_{\rm e}=0.3, \epsilon_{\rm B}=0.001, \gamma_{\rm max}=10^3, \gamma_{\rm min}=5, p=1.7$ (\textbf{black line}). For the equipartition case for PRS 121102, $E_{\rm 50}=0.7, R_{\rm 17}=3, \epsilon_{\rm e}=0.3, \epsilon_{\rm B}=0.3, \gamma_{\rm max}=10^3, \gamma_{\rm min}=5, p=1.7$ (\textbf{black dashed line}). For PRS 190520, $E_{\rm 50}=7.5, R_{\rm 17}=2.5, \epsilon_{\rm e}=0.3, \epsilon_{\rm B}=0.001, \gamma_{\rm max}=10^3, \gamma_{\rm min}=5, p=1.9$ (\textbf{black dotted line}). The results fit well with the observation data without the use of concrete models and details.
\label{fig:parameters estimation}}
\end{figure}

Finally we only have seven parameters, i.e. $E_{\rm 50},R_{\rm 17},\epsilon_{\rm B},\epsilon_{\rm e},\gamma_{\rm max},\gamma_{\rm min},p$, and take typical values to fit the spectrum of the PRS 121102 and PRS 190520.
Due to estimation in section \ref{sec:Parameter estimation}, we choose small $\epsilon_{\rm B}$ and $\epsilon_{\rm e}$ for there are other energy depletion of electron like gravitational and thermal loss, and appropriately low $\gamma_{\rm max}$ and $\gamma_{\rm min}$ for relativistic electrons in PRSs.
We fix $\gamma_{\rm max}=1000$ and $\gamma_{\rm min}=5$ hereafter for both PRSs due to their frequency ranges are nearly coincident. We also fix $\epsilon_{\rm e}=0.3$ and $\epsilon_{\rm B}=0.001$ except for equipartition case, for we consider the PRS is on its early age when the energy density of electron is greater than the magnetic energy density.
For PRS 121102, we set $E_{\rm 50}=3, R_{\rm 17}=1.5, \epsilon_{\rm e}=0.3, \epsilon_{\rm B}=0.001, \gamma_{\rm max}=10^3, \gamma_{\rm min}=5, p=1.7$ (\textbf{black line}), \textbf{and plot the spectrum in figure \ref{fig:parameters estimation}}. Note that for the observation of PRS 121102, $L_{\rm 1.7 GHz}\sim 2\times 10^{29}\rm \ erg \ s^{-1} \ Hz^{-1}$ and $L_{\rm 5 GHz}\sim 1.4\times 10^{29}\rm \ erg \ s^{-1} \ Hz^{-1}$.
For the equipartition case for PRS 121102 (i.e. the energy density of electron $\epsilon_{\rm e} U$ is equal to the magnetic energy density $\epsilon_{\rm B} U$), we set  $E_{\rm 50}=0.7, R_{\rm 17}=3, \epsilon_{\rm e}=0.3, \epsilon_{\rm B}=0.3, \gamma_{\rm max}=10^3, \gamma_{\rm min}=5, p=1.7$ (\textbf{black dashed line}). We find that the enhance of magnetic energy density would lead to a smaller $E_{\rm 50}$ and a larger $R_{\rm 17}$ to fit the observation, as well as giving rise to a smaller $\nu_{\rm a}$. As magnetic field enhances, the same distribution of electrons could emit at higher luminosity, thus requiring a smaller energy density $U$ to fit the observation. For PRS 190520, we set $E_{\rm 50}=7.5, R_{\rm 17}=2.5, \epsilon_{\rm e}=0.3, \epsilon_{\rm B}=0.001, \gamma_{\rm max}=10^3, \gamma_{\rm min}=5, p=1.9$ (\textbf{black dotted line}). Note that for the observation of PRS 190520, $L_{\rm 3 GHz}\sim 3.6\times 10^{29}\rm \ erg \ s^{-1} \ Hz^{-1}$ and the spectra index from 1.5 GHz to 5.5 GHz is $\sim -0.41$. The equipartition case for PRS 190520 is similar to that for PRS 121102, in which the enhance of magnetic energy density would lead to smaller energy density $U$ related to $E_{\rm 50}$ and $R_{\rm 17}$. We plot three cases together (see figure \ref{fig:parameters estimation}). 
The results fit well with the observation without the use of concrete models and details, and with more data we can determine the peak frequencies better, which shows that the synchrotron radiation of dense power-law electrons is account for the emission of PRSs. The fitting also shows that the intrinsic physical quantities such as energy injection and size for PRS 190520 is similar but larger than that for PRS 121102.

\subsection{Dynamical Evolution} \label{subsec:dynamical}

We consider a picture of a pulsar-like central engine, generating a cold ultra-relativistic wind propagating outwards. After collision and getting heated at the termination shock at $R_t$, a forward shock continues sweeping up the outside medium at $R_{\rm PWN}$ and a power-law distribution of electron is raised, giving rise to a PWN. \cite{2006ARA&A..44...17G} pointed out that $R_t$ is much smaller than $R_{\rm PWN}$, so we take the PWN as a spherical bulk rather than a bubble, and assume the dynamical evolution and radiation are in the same one zone.
The PWN is dynamical along with its radiation properties \citep{2017ApJ...838L...7D,2017ApJ...839L...3K,2017ApJ...841...14M,2018ApJ...868L...4M,2021ApJ...923L..17Z}.
Since we apply the dynamical PWN model on PRS, we can obtain any time-varying quantities as long as they are related to dynamical timescale.

\textbf{Now we discuss how the $R_{\rm PWN}$ evolves with time}, for a PWN sweeping up a freely expanding SNR \citep{2020Vink}. The contact discontinuity at about $R_{\rm PWN}$, is relatively thin so we consider it as a thin shell, which is dynamical due to the difference of the pressure and density at both sides. For conservation law of energy,
\begin{equation}\label{thermal}
    dQ=d(UV)+PdV,
\end{equation}
where $Q$ is the energy injection, $U$ is the internal energy density, $P$ is the pressure and $V$ is the bulk volume. For relativistic adiabatic gas, we have $U=\frac{P}{\gamma-1}$ and $\gamma=\frac{4}{3}$. Note that for relativistic gas, the sound speed $c_s=\frac{c}{\sqrt{3}}$, so the pressure is nearly uniform throughout the PWN. Taking the time derivative of (\ref{thermal}), and assume the energy injection rate $L_{\rm w}=\frac{dQ}{dt}$ is constant,
\begin{equation}
     L_{\rm w} =\frac{d(4\pi R^3_{\rm PWN} P_{\rm PWN})}{dt}+P_{\rm PWN}4\pi R^2_{\rm PWN}\frac{dR_{\rm PWN}}{dt}
=P_{\rm PWN} 16\pi R^2_{\rm PWN} \frac{dR_{\rm PWN}}{dt}+4\pi R^3_{\rm PWN}\frac{dP_{\rm PWN}}{dR_{\rm PWN}}\frac{dR_{\rm PWN}}{dt}.
\end{equation}
The motion of the contact discontinuity, neglecting the pressure of the unshocked ejecta, follows from
\begin{equation}
    M_{\rm cd}\frac{d^2 R_{\rm PWN}}{dt^2}=4\pi R^2_{\rm PWN} P_{\rm PWN},
\end{equation}
where $M_{\rm cd}=\frac{4}{3}\pi R^3_{\rm PWN} \rho_{\rm ej}$ is the swept-up mass of the ejecta. \textbf{We assume the ejecta is freely expanding, $\rho_{\rm ej}=\rho_0 (\frac{t}{t_0})^{-3}=n_0 m_{\rm p} (\frac{t}{t_0})^{-3}$}, where $m_{\rm p}$ is proton mass and $n_0$ is electron number density, and the characteristic time $t_0$ is set to be 1 yr as followed. We apply the self-similarity solution, $P_{\rm pwn}\propto R_{\rm PWN}^\alpha$ and $R_{\rm PWN}\propto t^{\rm m}$, from equation (24,25) we obtain
\begin{equation}
    m=\frac{6}{5},
\end{equation}
thus
\begin{equation}
    R_{\rm PWN}=0.71(\frac{L_{\rm w} t^6}{n_0 m_{\rm p}})^{\frac{1}{5}}=1.3\times 10^{18}(\frac{L_{\rm w,41} t_2^6}{n_{\rm 0,2}})^{\frac{1}{5}} \rm \ cm,
\end{equation}
\textbf{where the energy injection rate $L_{\rm w}=10^{41}L_{\rm w,41}\rm \ erg \ s^{-1}$, the age $t=10^2t_2\rm \ yr$, and the electron number density $n_0=10^2 n_{\rm 0,2} \rm \ cm^{\rm -3}$}.
Therefore, we can calculate the
total energy and energy density of the PWN,
\begin{equation}
    E_{\rm inj}=\frac{28\pi}{25} (0.71)^5 L_{\rm w} t=2.0\times 10^{50} L_{\rm w,41} t_2 \rm \ erg,
\end{equation}
\begin{equation}
    U=\frac{E_{\rm inj}}{\frac{4}{3}\pi R_{\rm PWN}^3}=2.3\times 10^{-5}L_{\rm w,41}^{\frac{2}{5}}n_{\rm 0,2}^{\frac{3}{5}}t_2^{-\frac{4}{5}} \rm \ erg \ cm^{-3}.
\end{equation}
Since cooling Lorentz factor $\gamma_{\rm c}$ and cooling frequency $\nu_{\rm c}$ are time-dependent, they are coupled to the dynamical evolution. By applying equation (14-22), we can obtain evolutionary spectra.
Note that for different physical phases or environments or mechanical analysis (e.g. different ages of PWN,  
different surroundings including SNR and interstellar medium, ram pressure), could have impact on the index $m$ of $R_{\rm PWN}$ (see section \ref{sec:discussion} for more detail).
In one zone model, DM and RM can be demonstrated approximately as \citep{2017ApJ...838L...7D,2019ApJ...885..149Y},
\begin{equation}
    DM \sim n_{\rm e} l \sim \frac{1}{p-1} K_{\rm e} \gamma_{\rm min}^{-p+1} R_{\rm PWN}\propto \frac{L_{\rm w,41}^{8/5-p}}{t_2^{7/5}}, 
\end{equation}
\begin{equation}
    RM \sim \frac{e^3}{2\pi m_{\rm e}^2 c^4} n_{\rm e} l B \sim \frac{e^3}{2\pi m_{\rm e}^2 c^4} \frac{1}{p-1} K_{\rm e} \gamma_{\rm min}^{-p+1} R_{\rm PWN} B \propto \frac{L_{\rm w,41}^{9/5-p}}{t_2^{27/10}}.
\end{equation}
As shown, both DM and RM depend more on $t_2$ than on $L_{\rm w,41}$, and the index of $t_2$ is free from $p$ which is the index of electron distribution.
In the next section, we can compare our PWN scenario with the observations of the PRS to constrain the model parameters.

\subsection{Parameter constraints and results} \label{sec:results}

\textbf{For our dynamical one-zone scenario of PWN, we have eight parameters: $L_{\rm w,41}, t_2, n_{\rm 0,2}, \epsilon_{\rm e}, \epsilon_{\rm B}, \gamma_{\rm max}, \gamma_{\rm min}, p$}. Note that for fixed value of $L_{\rm w,41}=13.4$ \textbf{(this value is set to fit the observation data of PRS 121102, as detailed below in this subsection)}, $n_{\rm 0,2}=1, \epsilon_{\rm e}=0.3, \epsilon_{\rm B}=0.001, \gamma_{\rm max}=10^3, \gamma_{\rm min}=5$, $\nu_{\rm c}$ is only a function of $t_2$,
\begin{equation}
    \nu_{\rm c}=8.8\times 10^{13} t^{\frac{19}{10}}_2 \rm \ Hz,
\end{equation}
where $\nu_{\rm c}$ increases with time as nearly $t^2$. \textbf{Here we assume $n_{\rm 0,2}$ to be 1 as a typical value as followed, to be consistent with that $\frac{\rho_{\rm ej}}{m_{\rm p}}\leq 1 \ \rm cm^{-3}$ in the following time evolution according to \cite{2022hxga.book...61M}}. For observational constraints, we obtain $t_2=0.0027$ by applying $\nu_{\rm c}=5.5\rm \ GHz$. From Extended Data Figure 2 of \cite{2017Natur.541...58C}, the spectral index segments at 10GHz, caused by either the segment of electrons’ distribution or the synchrotron cooling. Assuming $\nu_{\rm c}=10\rm \ GHz$, we obtain $t_2=0.0037$. Both indicate that in our scenario, synchrotron cooling frequency can increase significantly in the first few years or decades until the next phase of PWN and SNR, which makes $\nu_{\rm c}$ not a very good constraint on the parameters.

\begin{figure}[ht!]
\centering
\begin{minipage}{0.45\textwidth}
    \includegraphics[width=0.95\textwidth]{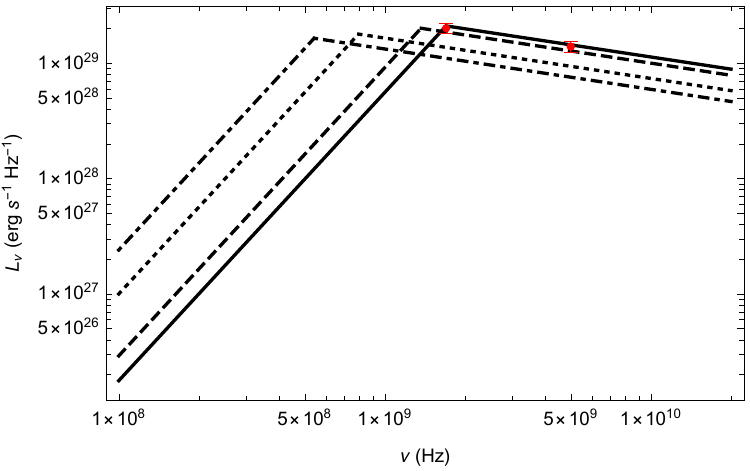}
\end{minipage}
\begin{minipage}{0.45\textwidth}
    \includegraphics[width=0.95\textwidth]{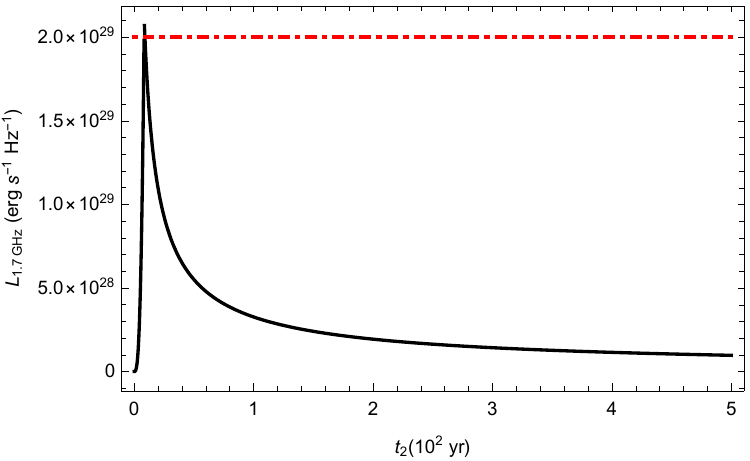}
\end{minipage}
\caption{We fix $L_{\rm w,41}=13.4, n_{\rm 0,2}=1, \epsilon_{\rm e}=0.3, \epsilon_{\rm B}=0.001, \gamma_{\rm max}=10^3, \gamma_{\rm min}=5, p=1.7$, set $t_2=0.085$ (\textbf{black}), $t_2=0.1$ (\textbf{black dashed}), $t_2=0.15$ (\textbf{black dotted}), $t_2=0.2$ (\textbf{black dot-dashed}), plot the spectrum of PRS 121102 at the left panel. The trend of the luminosity at 1.7 GHz is shown at the right panel, where the \textbf{red dot-dashed line} is the observation of $L_{\rm 1.7 GHz}\sim 2.0\times 10^{29}\rm \ erg \ s^{-1} \ Hz^{-1}$.
\label{fig:result and time}}
\end{figure}

We fix $L_{\rm w,41}=13.4, n_{\rm 0,2}=1, \epsilon_{\rm e}=0.3, \epsilon_{\rm B}=0.001, \gamma_{\rm max}=10^3, \gamma_{\rm min}=5, p=1.7$, set $t_2=0.085$ (\textbf{black}), $t_2=0.1$ (\textbf{black dashed}), $t_2=0.15$ (\textbf{black dotted}), $t_2=0.2$ (\textbf{black dot-dashed}), \textbf{and plot the spectrum of PRS 121102 (see the left panel in figure \ref{fig:result and time})}.
Via adjusting age $t_2$ and spectrum index $p$, we can fit the following observation well in future.
\textbf{We also plot the luminosity of PRS 121102 at 1.7 GHz in the $t_2$ range of 0.01$\sim$5 (see the right panel in figure \ref{fig:result and time})}. The luminosity falls rapidly in the first 100 years, then becomes shallow with approximation to \textbf{about $10^{28}\rm \ erg \ s^{-1}$ in the following 400 years}. Note that since the thermal depletion and radiation depletion of PWN is neglectable and $\nu_{\rm c}$ is increasing, the synchrotron platform is steady relatively, which indicates luminosity at other frequency should follow the same trend in our scenario.

\begin{figure}
    \centering
    \begin{minipage}{0.45\textwidth}
        \includegraphics[width=0.95\textwidth]{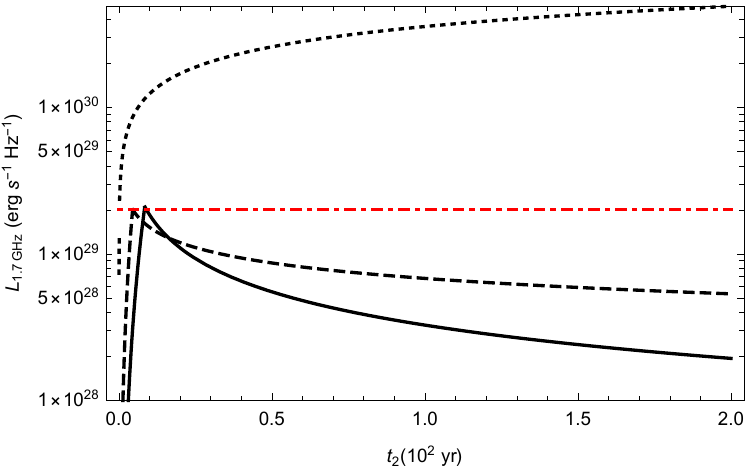}
    \end{minipage}
    \begin{minipage}{0.45\textwidth}
        \includegraphics[width=0.95\textwidth]{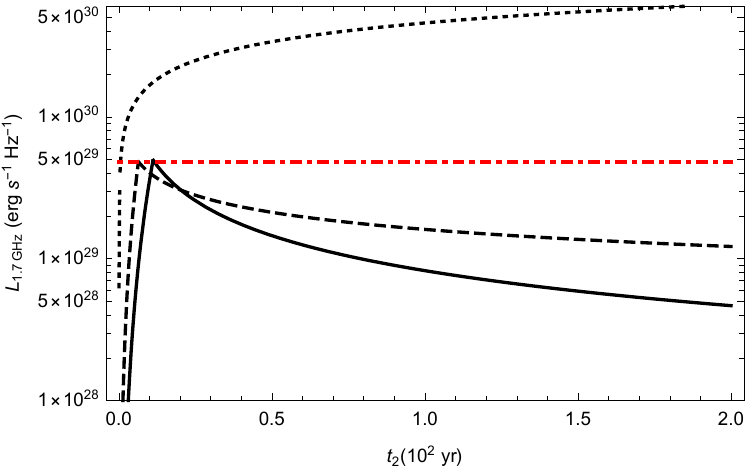}
    \end{minipage}
   
    \caption{For PRS 121102 in the left panel, the black line is $R_{\rm PWN}\propto t_2^{\frac{6}{5}}$, the \textbf{black dashed line} is $R_{\rm PWN}\propto t_2$, the \textbf{black dotted line} is $R_{\rm PWN}\propto t_2^{\frac{3}{5}}$, the parameters are taken that  $L_{\rm w,41}=13.4$ for $R_{\rm PWN}\propto t_2^{\frac{6}{5}}$, $L_{\rm w,41}=23.8$ for $R_{\rm PWN}\propto t_2$, $L_{\rm w,41}=537.4$ for $R_{\rm PWN}\propto t_2^{\frac{3}{5}}$, $n_{\rm 0,2}=1, \epsilon_{\rm e}=0.3, \epsilon_{\rm B}=0.001, \gamma_{\rm max}=10^3, \gamma_{\rm min}=5, p=1.7$, where the \textbf{red dot-dashed line} is the observation of $L_{\rm 1.7 GHz}\sim 2.0\times 10^{29}\rm \ erg \ s^{-1} \ Hz^{-1}$. For PRS 190520 in the right panel, the black line is for $R_{\rm PWN}\propto t_2^{\frac{6}{5}}$, the \textbf{black dashed line} is for $R_{\rm PWN}\propto t_2$, the \textbf{black dotted line} is for $R_{\rm PWN}\propto t_2^{\frac{3}{5}}$. The parameters are taken that  $L_{\rm w,41}=22.6$ for $R_{\rm PWN}\propto t_2^{\frac{6}{5}}$, $L_{\rm w,41}=38.1$ for $R_{\rm PWN}\propto t_2$, $L_{\rm w,41}=521.2$ for $R_{\rm PWN}\propto t_2^{\frac{3}{5}}$, $n_{\rm 0,2}=1, \epsilon_{\rm e}=0.3, \epsilon_{\rm B}=0.001, \gamma_{\rm max}=10^3, \gamma_{\rm min}=5, p=1.8$. The \textbf{red dot-dashed line} is the observation of $L_{\rm 1.7 GHz}\sim 4.5\times 10^{29}\rm \ erg \ s^{-1} \ Hz^{-1}$. The comparison of both PRSs shows the spectra of two PRSs have similar trends, neglecting the different magnitude of luminosity.}
    \label{fig:comparison2}
\end{figure}

To compare differences of both PRSs, we plot the luminosity of PRS 190520 at 1.7 GHz in the $t_2$ \textbf{range of 0.001$\sim$2} (see the right panel in figure \ref{fig:comparison2}), under three different phases, i.e. $R_{\rm PWN}\propto t_2^{\frac{6}{5}}$ (black line), $R_{\rm PWN}\propto t_2$ (\textbf{black dashed line}), $R_{\rm PWN}\propto t_2^{\frac{3}{5}}$ (\textbf{black dotted line}). Each phase indicate a different physical state, e.g. a PWN embedded in a freely expanding SNR for $R_{\rm PWN}\propto t_2^{\frac{6}{5}}$, a PWN embedded in a stellar wind with constant flux for $R_{\rm PWN}\propto t_2$ \citep{2019ApJ...885..149Y} and a PWN embedded in static medium for $R_{\rm PWN}\propto t_2^{\frac{3}{5}}$ \citep{2017ApJ...838L...7D}. We show that $R_{\rm PWN}\propto t_2^{\frac{3}{5}}$, i.e. the surrounding is nearly static with respect to PWN, demonstrates a different trend compared to other two phases, which is increasing rather than decreasing in a long period. 
For fixed parameters $n_{\rm 0,2}=1, \epsilon_{\rm e}=0.3, \epsilon_{\rm B}=0.001, \gamma_{\rm max}=10^3, \gamma_{\rm min}=5, p=1.8$, $L_{\rm w,41}$ is taken to be 22.6 for $R_{\rm PWN}\propto t_2^{\frac{6}{5}}$, 38.1 for $R_{\rm PWN}\propto t_2$, 521.2 for $R_{\rm PWN}\propto t_2^{\frac{3}{5}}$, to be insure that $\nu_{\rm a}$ is 1.5 GHz and $L_{\rm 1.5\rm GHz}$ is $4.8\times 10^{29}\rm \ erg \ s^{-1} \ Hz^{-1}$ for which are consistent with the observation by now.
We plot the luminosity of PRS 121102 at 1.7 GHz for comparison (see the left panel in figure \ref{fig:comparison2}), since the luminosity at 1.7 GHz is the observation data of PRS 121102. It is shown that both PRSs are very similar except for the intrinsic energy. It seems that if radiation of PRS varies significantly, it is more likely due to different phases which PWN is going through.

\begin{figure}[ht!]
\centering
\begin{minipage}{0.45\textwidth}
    \includegraphics[width=0.95\textwidth]{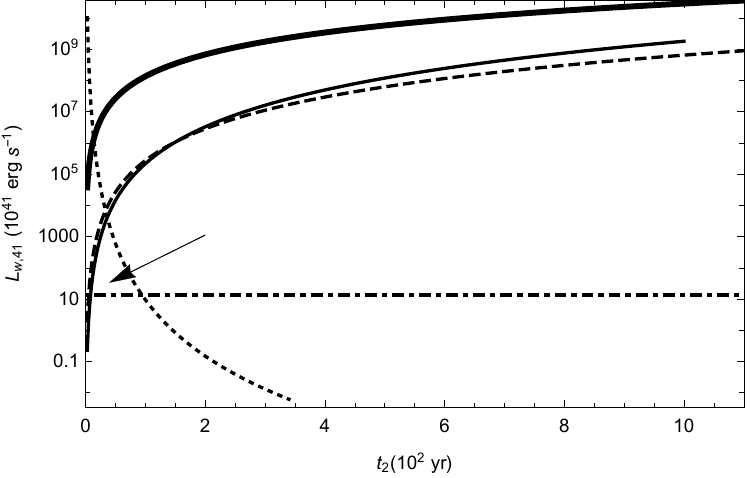}
\end{minipage}
\begin{minipage}{0.45\textwidth}
    \includegraphics[width=0.95\textwidth]{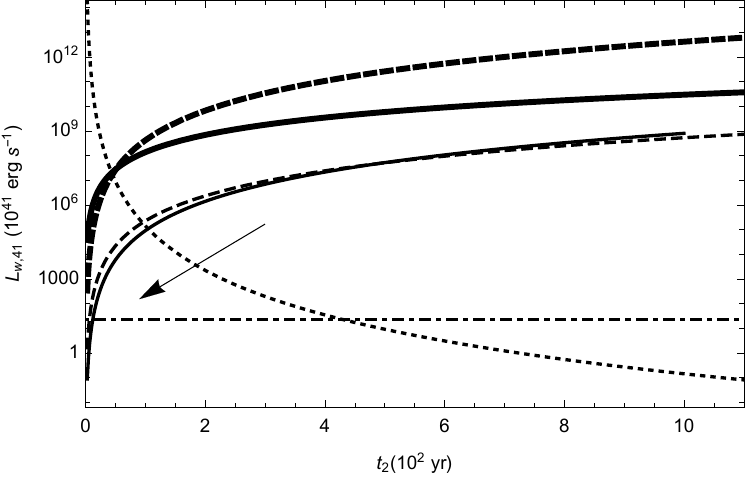}
\end{minipage}
\caption{Limits \textbf{(1) $\nu_{\rm a}=1.7\rm \ GHz$ (black line), (2) $L_{\rm 1.7\rm GHz}\approx 2\times10^{29}\rm \ erg \ s^{-1} \ Hz^{-1}$ (\textbf{black dot-dashed line}), (3) the upper limit of PWN size $\sim 0.66\rm \ pc$ (\textbf{black dotted line}), (4) DM $\sim565.8 \rm \ pc cm^{-3}$ (\textbf{black thick line}), (5) RM $\sim 10^5 \rm \ rad \ m^{-2}$ (\textbf{black dashed line})} for PRS 121102 in the left panel. Limits \textbf{(1) $\nu_{\rm a}\sim 1.5\rm \ GHz$ (black line), (2) $L_{\rm 1.5\rm GHz}\sim 4.8\times 10^{29}\rm \ erg \ s^{-1} \ Hz^{-1}$ (\textbf{black dot-dashed line}), (3) the upper limit of PWN size $\sim 4.5\rm \ pc$ (\textbf{black dotted line}), (4) DM $\sim 1204.7\rm \ pc \ cm^{-3}$ (\textbf{black thick line}), (5) RM $\sim 1.8\times10^5 \rm \ rad \ m^{-2}$ (\textbf{black dashed line}), (6) decrease of DM $\sim 0.09\rm \ pc \ cm^{-3} \ day^{-1}$ (\textbf{black thick dashed line})} for PRS 190520 in the right panel. Both permitted parameter spaces of $L_{\rm w,41}$ and $t_2$ are pointed out with black arrow. \textbf{DM constraints from both panels do not constrain the permitted parameter space any better than the other parameters}.
\label{fig:ranges}}
\end{figure}

We fix  $n_{\rm 0,2}=1, \epsilon_{\rm e}=0.3, \epsilon_{\rm B}=0.001, \gamma_{\rm max}=10^3, \gamma_{\rm min}=5, p=1.7$ for PRS 121102, and obtain the constraints on the source parameters $L_{\rm w,41}$ and $t_2$ through \textbf{(1) $\nu_{\rm a}$ (black line), (2) $L_{\rm 1.7\rm GHz}$ (\textbf{black dot-dashed line}), (3) the upper limit of PWN size (\textbf{black dotted line}), (4) DM (\textbf{black thick line}), (5) RM (\textbf{black dashed line}) (see figure \ref{fig:ranges}).} The $\nu_{\rm a}$ decreases with $t_2$ but increases with $L_{\rm w,41}$, and it must be $\leq1.7\rm \ GHz$ by now to be consistent with the observation. We draw the black line for this constraint that any $L_{\rm w,41}$ and $t_2$ satisfy $\nu_{\rm a}=1.7\rm \ GHz$, where the region below the black line is the permitted parameter space. The second constraint is that $L_{\rm 1.7\rm GHz}\approx 2\times10^{29}\rm \ erg \ s^{-1} \ Hz^{-1}$ by now. As $\nu_{\rm a}$ and the luminosity of $\nu_{\rm a}$ (the peak luminosity) decrease with time, a smaller $\nu_{\rm a}$ requires a larger $L_{\rm w,41}$ to maintain $L_{\rm 1.7\rm GHz}\approx 2\times10^{29}\rm \ erg \ s^{-1}$, which indicates that a largest $\nu_{\rm a}$, i.e. $\nu_{\rm a}=1.7\rm \ GHz$, will suggest a smallest $L_{\rm w,41}$. We work out $L_{\rm w,41}\ge 13.4$ to satisfy this constraint. $R_{\rm PWN}$ is a function only of $L_{\rm w,41}$ and $t_2$ and the observation limit is 0.66 pc. We draw the \textbf{black dotted line} for the size constraint. Both DM and RM are functions only of $L_{\rm w,41}$ and $t_2$, and the observation limits are $565.8 \rm \ pc \ cm^{-3}$ and $\sim 10^5 \rm \ rad \ m^{-2}$. Here we assume that DM and RM are originated from the electron and magnetic field in the one-zone instead of outside surroundings. The final permitted parameter space is surrounded by five lines. The source ages between about 8.5$\sim$100 yr, which is consistent with the assumption of our scenario that the whole system is at its early age, when the thermal depletion and radiation depletion of PWN can be neglectable and the unshocked ejecta is still freely expanding. Meanwhile, the energy injection rate ranges from about $10^{42}\rm \ erg \ s^{-1}$ up to $10^{44}\rm \ erg \ s^{-1}$, which might indicate that the central engine is significantly energetic at its early age, more likely leading to the magnetic energy budget of magnetar. Note that the $\nu_{\rm a}$ limit fits well with the RM limit, while they are all far from the DM limit, which may suggest that for such circumstance the DM value should be supplied by other ``zones''.

We fix $n_{\rm 0,2}=1, \epsilon_{\rm e}=0.3, \epsilon_{\rm B}=0.001, \gamma_{\rm max}=10^3, \gamma_{\rm min}=5, p=1.8$ for PRS 190520 and apply the same constraints as \textbf{(1) $\nu_{\rm a}\sim 1.5\rm \ GHz$ (black line), (2) $L_{\rm 1.5\rm GHz}\sim 4.8\times 10^{29}\rm \ erg s^{-1} Hz^{-1}$ (\textbf{black dot-dashed line}), (3) the upper limit of PWN size $\sim 4.5\rm \ pc$ (\textbf{black dotted line}), (4) DM $\sim 1204.7\rm \ pc \ cm^{-3}$ (\textbf{black thick line}), (5) RM $\sim 1.8\times10^5 \rm \ rad \ m^{-2}$ (\textbf{black dashed line}), (6) decrease of DM $\sim 0.09\rm \ pc \ cm^{-3} \ day^{-1}$ (\textbf{black thick dashed line}) (see figure \ref{fig:ranges}).} It seems that the allowed parameter regime for PRS 190520 is broader than that for PRS 121102, i.e. for PRS 190520 the energy injection rate ranges from about $10^{42}\rm \ erg \ s^{-1}$ up to $10^{46}\rm \ erg \ s^{-1}$, and the age ranges from about 12.3$\sim$400 yr. Similar to PRS 121102, the main constraints come from $\nu_{\rm a}$, $L_{\rm 1.5\rm GHz}$, size and RM. Meanwhile, the additional limit of decrease of DM does not meet the limit of DM itself for long, which may lead to that the generating and the decreasing of DM could originate from different ``zones''.

\section{Discussion} \label{sec:discussion}
By applying one-zone model to the PRS as a PWN, we find that $R_{\rm PWN}$, as a function of time, is the most characteristic physical quantity for different systems. \cite{2017ApJ...838L...7D} proposed a PWN with static ambient medium, which obtains $R_{\rm PWN}\propto t^{\frac{3}{5}}$; \cite{2018ApJ...868L...4M} proposed a PWN that is freely expanding, which obtains $R_{\rm PWN}\propto t$; \cite{2019ApJ...885..149Y} proposed a PWN embedded in a solar wind with a constant flux, which also obtains $R_{\rm PWN}\propto t$; in this paper, we propose a PWN with freely expanding SNR and obtain $R_{\rm PWN}\propto t^{\frac{6}{5}}$. The surroundings outside the PWN impact the time evolution of $R_{\rm PWN}$, mainly because of the ram pressure on the PWN; as the surroundings are able to transit to different phases, e.g. a SNR freely expands at early time and nearly becomes the static ambient medium at the end, the PWN is expected to have different phases similarly, i.e. a PWN can go through three phases above at different age. \cite{2021ApJ...923L..17Z}proposed a PWN that goes through two phases before and after the PWN catches up with the SNR. Yet a unified model of PWN including different phases is incomplete, and also the thermal depletion and radiation depletion of both the PWN and the surroundings should be considered since they might trigger the phase transition. However, as highly dependent on different phases, the index m of $t^{\rm m}$ in the function of $R_{\rm PWN}$ \textbf{is a good parameter worth continued monitoring through observations, which might be a probe} into the phases transition and physical environment of the surroundings. Another important physical quantity is the energy injection rate $L_{\rm w}$, i.e. the function form of $L_{\rm w}$. In \cite{2017ApJ...838L...7D,2019ApJ...885..149Y} and our paper, $L_{\rm w}$ is assumed to be a constant; in \cite{2018ApJ...868L...4M,2021ApJ...923L..17Z}, $L_{\rm w}$ is an power function of time; alternatively, $L_{\rm w}$ can be a pulse function probably, e.g. the FRB bursts \citep{2020ApJ...896...71L}, in which the PWN might freely expand like the SNR. Different kinds of $L_{\rm w}$ function can make a difference to the dependence of $R_{\rm PWN}$ on time at every phase of PWN.

\begin{figure}
    \centering
    \includegraphics{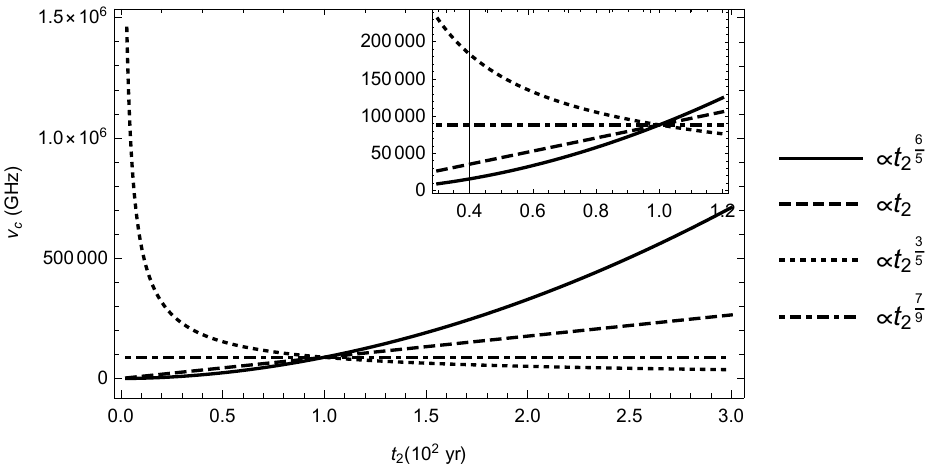}
    \caption{Different trend of $\nu_{\rm c}$ under different index $m$ for PRS 121102. Four lines confluent at $t_2=1$.}
    \label{fig:cooling freq.}
\end{figure}

$\nu_{\rm c}$ is a sensitive physical quantity to these phases. As shown in figure \ref{fig:cooling freq.}, for fixed values $L_{\rm w,41}=13.4, n_{\rm 0,2}=1, \epsilon_{\rm e}=0.3, \epsilon_{\rm B}=0.001, \gamma_{\rm max}=10^3, \gamma_{\rm min}=5, p=1.7$, different index $m$ in $R_{\rm PWN}\propto t^{\rm m}$ impacts the evolution of $\nu_{\rm c}$. We find that for $R_{\rm PWN}\propto t^{\frac{3}{5}}$, the $\nu_{\rm c}$ is decreasing, while it is increasing for $R_{\rm PWN}\propto t$ and $R_{\rm PWN}\propto t^{\frac{6}{5}}$, and we obtain that $\nu_{\rm c}$ does not change for $R_{\rm PWN}\propto t^{\frac{7}{9}}$ as the critical state.

For fixed values of  $n_{\rm 0,2}, \epsilon_{\rm e}, \epsilon_{\rm B}, \gamma_{\rm max}, \gamma_{\rm min}, p$, we can obtain the energy injection rate $L_{\rm w,41}$ and the age of the source $t_2$ through comparison with observation. However, for many FRBs and their potential PRSs, if their RMs $\ll 10^5\rm \ rad \  m^{-2}$, with $t_2$ ranging from about 0$\sim$1, their $L_{\rm w,41}$ should be $\ll 1$, which implies that their RMs should be much larger than observation for now to be consistent with the above permitted parameter region of two PRSs otherwise it is not likely to see these PRSs in GHz band. We suppose that if the permitted parameter region is solid for the PWN systems as PRSs, these systems would generate enough relativistic electrons to produce large RM while others would not, and the dense electron gas or the shock may create turbulence to depolarize the electromagnetic wave \citep{2022ApJ...928L..16Y}. This is the case for both PRS 121102 and PRS 190520 so their constraints and permitted parameter regions are similar. We also calculate the inverse-Compton scattering peak luminosity scattered by different $\gamma$ factor,
\begin{equation}
    \frac{\nu_{\rm ICS,peak}L_{\rm ICS,peak}}{\nu_{\rm syn,peak}L_{\rm syn,peak}}=\frac{U_{\rm ph}}{U_{\rm B}},
\end{equation}
where $U_{\rm ph}=\frac{\nu_{\rm syn,peak}L_{\rm syn,peak}}{4\pi R^2 c}$ is radiation field energy density, and $\nu_{\rm ICS,peak}=\gamma^2\nu_{\rm syn,peak}$.We assume $\nu_{\rm syn,peak}$ is 1.7 GHz for PRS 121102 and 1.5 GHz for PRS 190520 as observation, $R$ is 0.66 pc for PRS 121102 and 4.5 pc for PRS 190520. Though ICS peak frequency and luminosity are highly dependent on the parameter value, the luminosity is low as we may not be able to observe, e.g. for $\nu=10^{14} \rm \ Hz$, $L_{\rm \nu}\sim 2.2\times 10^{19}\rm \ erg \ s^{-1} \ Hz^{-1}$ for PRS 121102 and $L_{\rm \nu}\sim 4.0\times 10^{18}\rm \ erg \ s^{-1} \ Hz^{-1}$ for PRS 190520. Another possibility for high energy emission was proposed by \cite{2023ApJ...954..154C} that short gamma ray burst comes after an FRB-like burst produced from precursor wind in binary merger. Due to small source sample, we will need more observation data to determine the more appropriate typical values of all parameters in our scenario.

\textbf{For the third PRS recently discovered to be associated with FRB 201124A \citep{2023arXiv231215296B}, which needs further confirmation and observation, our model still can explain its high luminosity practically. We set $E_{\rm 50}=1.2, R_{\rm 17}=2.5, \epsilon_{\rm e}=0.3, \epsilon_{\rm B}=0.001, \gamma_{\rm max}=10^3, \gamma_{\rm min}=5, p=1.7$, in which $p$ is set to be 1.7 for comparison with PRS 121102, then we can obtain $L_{\rm 22 GHz}\approx 6.1\times 10^{\rm 27}\rm \ erg \ s^{-1} \ Hz^{-1}$. For the rest observation data at 15 Ghz and 6 GHz, we set $E_{\rm 50}=0.9, R_{\rm 17}=2.5, \epsilon_{\rm e}=0.3, \epsilon_{\rm B}=0.001, \gamma_{\rm max}=10^3, \gamma_{\rm min}=5, p=1.7$ to obtain $L_{\rm 15 GHz}\approx 4.1\times 10^{\rm 27}\rm \ erg \ s^{-1} \ Hz^{-1}$, $E_{\rm 50}=0.5, R_{\rm 17}=2.8, \epsilon_{\rm e}=0.3, \epsilon_{\rm B}=0.001, \gamma_{\rm max}=10^3, \gamma_{\rm min}=5, p=1.7$ to obtain $L_{\rm 6 GHz}\approx 1.7\times 10^{\rm 27}\rm \ erg \ s^{-1} \ Hz^{-1}$. Compared to PRS 121102, PRS 201124 has lower luminosity, DM and RM thus leading to a smaller $E_{\rm 50}$ and a bigger $R_{\rm 17}$ respectively, which is consistent with our scenario. As for the inverted spectrum index, which differs from the former two PRSs, we consider that there should be extra mechanism to explain it since the acceleration of shock wave often forms a power-law distribution of electron. Here we propose a few possible mechanism which may form the inverted spectrum index, e.g. the free-free absorption or Razin effect of plasma, the self-absorption or collision of two different distributions of electron and inhomogeneity of the source.}

One-zone model is a practical and most simple method to obtain properties, e.g. the spectra or evolution, for PRSs. 
It can demonstrate the constraints via observation and test the relation between parameterized physical quantities. With simple parameters, the energy injection $E_{\rm inj}$ and a size $R_{\rm PWN}$, we can obtain the hard spectrum of PRSs; when considering the dynamical evolution, the major physical quantity is $R_{\rm PWN}$ as a function of $t_2$ and energy injection rate $L_{\rm w,41}$, with which we are able to obtain time-varying of relevant properties.
This parameterized model does not depend on concrete physical models such as central engine or other details; instead via comparison of parameters and observation, we can constraint physical model, e.g. the dynamical evolution greatly involves with the index $m$ of $R_{\rm PWN}\propto t^{\rm m}$, where different $m$ indicates different phases and surroundings. The same case is also capable for energy injection rate $L_{\rm w}$. In our scenario, we only assume a PWN with magnetic field, similar to a energy-driven afterglow of gamma ray burst, in which a synchrotron emission is raised by a power-law distribution of electrons. For special case, we calculate $m=\frac{6}{5}$ and point out its phase that a PWN embedded in a freely expanding SNR neglectable of thermal and radiation loss in \ref{subsec:dynamical}, and show the permitted parameter space consistent with the assumption of early age of PWN through nowaday observation in \ref{sec:results}. 

We propose such a generic model, only with basic energy injection and electron distribution (i.e. a PWN assumption), to describe all properties of PRS, in which we can add more details like electron cooling or radiation loss. This model can also check the relation between properties and if they are truly in one zone or not, e.g. if the permitted parameter space is valid, then the DM shoule be smaller than observation thus indicating another zone to generate DM. However, with nowaday observation, the constraints are not enough to well determine the phase or surroundings, and there can be other zones or physical processes like inhomogeneous expanding which are beyond one-zone model.
With more potential PRSs and more monitoring observation in the future, we can further determine the phases of PRSs themselves and get more implication of central engine.

\begin{acknowledgments}
    This work is supported by National SKA Program of China
(2020SKA0120300) and NSFC (11773008).
\end{acknowledgments}

\bibliography{sample631}{}
\bibliographystyle{aasjournal}


\end{CJK}
\end{document}